\documentclass{jpsj-suppl}
\usepackage{txfonts} 

\usepackage{graphicx}        
\newcommand{\be}{\begin{equation}}
\newcommand{\ee}{\end{equation}}
\newcommand{\bra}{\langle}
\newcommand{\ket}{\rangle}
\newcommand{\bea}{\begin{eqnarray}}
\newcommand{\eea}{\end{eqnarray}}
\newcommand{\dis}{\displaystyle}

\title{Realized Volatility Analysis in A Spin Model of Financial Markets}

\author{Tetsuya \textsc{Takaishi}$^{1}$ }

\inst{$^{1}$Hiroshima University of Economics, 5-37-1, Gion, Asaminami-ku , Hiroshima 731-0192 , Japan}

\email{tt-taka@hue.ac.jp}

\abst{
We calculate the realized volatility in the spin model of financial markets
and examine the returns standardized by the realized volatility.
We find that moments of the standardized returns 
agree with the theoretical values of standard normal variables.
This is the first evidence that the return dynamics of the spin financial market is consistent 
with the view of the mixture-of-distribution hypothesis that also holds in the real financial markets.
}

\kword{Volatility, Mixture-of-distribution hypothesis, Spin model, Financial Markets}

\begin{document}
\maketitle

\vspace{-1.2cm}

\section{Introduction}
It is well-known that  
asset price time series show universal properties that are not
observed in a random walk model:
Price returns show fat-tailed distributions. 
Autocorrelations of returns are not
significant. On the other hand absolute returns are long correlated.
Volatility clustering occurs, etc.
These properties are now classified as stylized facts of asset returns\cite{CONT}.
A possible explanation for these properties  has been given by Clark\cite{Clark}
who suggested that the return dynamics follows the Gaussian random process with time-varying volatility,
called the mixture-of-distribution hypothesis (MDH).
Under the MDH, each return at time $t$ is described by $r_t=\sigma_t \epsilon_t$,
where $\sigma_t^2$ is the time-varying volatility and $\epsilon_t$ is a standard normal value $\sim N(0,1)$.
The return distributions from the MDH are given as a superposition of the volatility distribution and 
the conditional Gaussian distribution. Empirically the volatility distribution is suggested to be 
the inverse gamma distribution with which the unconditional return distribution 
results in the Student's t-distribution\cite{Super}. 
The MDH can be verified by examining the returns standardized by $\sigma_t$.
If the MDH holds, 
the standardized returns (SR) are given by $\bar{r}_t =r_t/\sigma_t=\epsilon_t$
and thus we should observe the normality for  $\bar{r}_t$, e.g. kurtosis=3.
A drawback of this verification is that volatility is not directly measurable in financial markets.

Recent availability of high-frequency intraday returns enables us to construct realized volatility (RV)\cite{RV,RV3}
that 
converges to the true volatility as the sampling interval goes to zero.  
Using the RV the normality of the SR in the financial markets has been studied\cite{RV3,RV4,RV6,Takaishi} and
it is shown that the SR are approximately described by normal variables, which supports the view of the MDH.

To better understand the origin of the price dynamics observed in the financial markets 
Bornholdt proposed a mimimalistic spin model that includes only two conflicting interactions\cite{Born}.
One of the interactions corresponds to the majority effect that agents imitate their neighbors. The other is the effect that 
agents tend to join minority groups.
The model with these conflicting interactions shows a non-equilibrium dynamics in return time series and 
exhibits major stylized facts successfully\cite{Born,Born2,POTTS,Born3}.
What has not yet been examined for the model is the consistency check of the MDH.
In this study we perform simulations of the spin model by Bornholdt
and examine if the MDH also holds for the spin model. 
Following the check process of the MDH in the real financial markets,
first we calculate the RV and then examine the normality of returns standardized by the RV.
We also examine higher moments of the SR that have not been used in the normality check  
in the real financial markets.
As seen later, the higher moments of the SR also support the MDH.

\section{Spin Financial Market}
The spin financial market by Bornholdt\cite{Born} is as follows.
We take an $L\times L$ lattice which has $N=L\times L$ sites.
Each site of the lattice has a spin agent $s_i$ which takes $+1$ or $-1$.
$s_i=+1$ $(-1)$ state can be assigned to "Buy" ("Sell") state.
Each spin $s_i(T)$ at time $T$ is updated to $s_i(T+1)$ 
according to the following probability by the heat-bath dynamics\cite{Born}.
\bea
s_i(T+1) & =   +1  &  p=1/(1+\exp(-2\beta h_i(T))), \\ \nonumber
s_i(T+1) & =  -1    & 1- p,
\label{eq:Prob}
\eea
where $h_i(T)$ is given by $\dis h_i(T)= \sum_{\bra i,j\ket} J s_j(T) -\alpha s_i(T)|M(T)|$,
$\bra i,j \ket$ stands for summation over the nearest neighbor pairs,  
and $M(T)$ is the magnetization given by $M(T)=\frac1N\sum_{i=1}^N s_i(T)$.
Here the unit of time $T$ is one sweep that means $N$ spins are updated.
The first term of $\dis h_i(T)$ corresponds to the nearest neighbor interaction of the ordinal Ising model 
that causes the majority effect with a positive $J$, i.e. agents imitate their neighbors.
The second term of $\dis h_i(T)$ corresponds to the minority effect 
that agents tend to belong to a minority group.   
$M(T)$ which represents the asymmetry of Buy and Sell states can be related to the price movements.
We define the market price $\ln p(T)$ by the fundamental price $p^*(T)$ and $M(T)$ 
as $\dis \ln p(T) = \ln p^*(T) +\lambda M(T)$ \cite{Born2}.
$\lambda$ can fix the magnitude of the log-price that depends on the actual asset price.
Since the value of $\lambda$ is not important for RV calculations, we set $\lambda=0.5$.
Assuming that the fundamental price is constant over time,
the price return $R(T)$, i.e. log-price difference is given
as \cite{Born2}
$
\dis
R(T)=\ln p(T+1)-\ln p(T)=(M(T+1)-M(T))/2. 
$

We define "one day" by one sweep.
Then $R(T)$ can be assigned to a daily return of the spin financial market.
We define the intraday time unit $t$ as one spin update.
Thus one day consists of $N$ updates.
The RV is constructed by a sum of squared intraday returns\cite{RV,RV3}.
For sampling interval $\Delta t$ we obtain $n+1$ intraday log-prices on each day as 
$(\ln p(T),\ln p(T+\Delta t/N),\ln (T+2\Delta t/N),...,\ln p(T+1))$, where $n=N/\Delta t$.
The $l$-th intraday return with $\Delta t$ on day $T$ is defined by
$\dis 
ret_{T,\Delta t}(l)=\ln p(T+l\Delta t/N)-\ln p(T+(l-1)\Delta t/N)
= (M(T+l\Delta t/N)-M(T+(l-1)\Delta t/N))/2,
$
Using $n$ intraday returns  the RV on $T$ with $\Delta t$ is defined by
\be
RV_{T,\Delta t}=\sum_{l=1}^{n=N/\Delta t} ret_{T,\Delta t}^2(l).
\label{eq:RV}
\ee

Let us assume that the log-price process follows a continuous stochastic diffusion,
$\dis d\ln p(s)=\tilde{\sigma}(s)dW(s)$, where $W(s)$ stands for a standard Brownian motion
and $\tilde{\sigma}(s)$ is a spot volatility at time $s$ that is not directly observed in the markets.
For this process, one-day integrated volatility is defined by
\be
\sigma^2(T) =\int_T^{T+1} \tilde{\sigma}^2(s) ds.
\label{eq:INT}
\ee
Eq.(\ref{eq:RV}) is a discretized version of eq.(\ref{eq:INT}) that goes to 
the true integrated volatility in the limit of $\Delta t \rightarrow 0$ when no bias exists.
Assuming the MDH for the daily return dynamics, 
we obtain $\dis R(T) =\sigma(T) \epsilon_T$.
Using the RV as $\sigma(T)^2$
the standardized return is given by $ \dis \bar{R}(T)= R(T)/RV_{T,\Delta t}^{1/2}$
that is used for the normality check.

\section{Finite-Sample Effect}
The RV by eq.(\ref{eq:RV}) depends on $\Delta t$ and
the true volatility is obtained only in the limit of $\Delta t \rightarrow 0$.
The distribution of the SR at finite $\Delta t$ is theoretically known 
and depends on the number of samples rather than $\Delta t$.
The distribution of the SR is given by\cite{Finite}
\be
f(\bar{R})=\frac{\Gamma(n/2)}{\sqrt{\pi n}\Gamma((n-1)/2)}\left(1-\frac{\bar{R}^2}{n}\right)^{(n-3)/2}
\times I(-\sqrt{n} \leq \bar{R} \leq \sqrt{n}),
\label{eq:finite}
\ee
where $n=N/\Delta t$ and 
$I(A)$ stands for the indicator function, i.e. $I(A)=1$ if $A$ is true and $I(A)=0$ if $A$ is false.
Under eq.(\ref{eq:finite}) the even moments $m^{2k}$ of $\bar{R}$ are calculated to be
$
\dis
m^{2k} =\frac{n^k (2k-1)(2k-3)\dots 1}{(n+2k-2)(n+2k-4)\dots n} 
$
that depends on $n$ except for variance $m^2$.
Empirically the finite-sample effect on the kurtosis of the SR has been observed 
in the real financial markets\cite{Andersen1,Andersen2,Takaishi}.

\section{Simulation Study}

Our simulations are performed on a 125$\times$125 square lattice with the periodic boundary condition.
The simulation parameters are set to $(\beta,\alpha,J)=(1.8,22,1)$.
Spins are randomly updated according to eq.(\ref{eq:Prob}).
We start the simulations on a lattice with ordered spins and discard the first $5\times10^3$ sweeps
as thermalization.
Then we collect data from $3\times10^4$ sweeps for analysis. 

\begin{figure}
\vspace{-1mm}
\centering
\includegraphics[height=4.0cm,width=11cm]{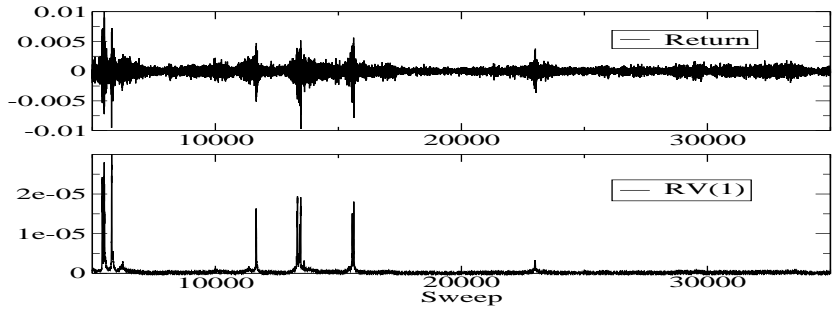}
\caption{
Sweep history of return and RV($\Delta t=1$). 
}
\vspace{-6mm}
\label{fig:History}
\end{figure}

Fig.1 shows returns at each sweep $(T)$ and their corresponding RV's calculated at $\Delta t=1$.
As seen in the figure the return time series shows a non-Gaussian process.
To verify the MDH we standardize each return by the RV calculated at various sampling intervals
$(\Delta t=1,2,...,4000)$.
and examine moments of the SR.
Fig.2 shows kurtosis, 6th, 8th and 10th moments of the SR as a function of $\Delta t$. 
They strongly depend on  $\Delta t$ 
and deviate from the expected values for normal variables at large $\Delta t$. 
However they approach the corresponding theoretical values as $\Delta t$ decreases.
To obtain the values at $\Delta t \rightarrow 0$ we fit them to the expected theoretical curves with one free parameter, i.e. 
$2k$-th moment is fitted to the function of $\dis \frac{C n^k }{(n+2k-2)(n+2k-4)\dots n}$
with a  parameter $C$.
Table I shows the fitting results which correspond to the values at $\Delta t \rightarrow 0$.
We find that the fitting results are very close to the theoretical results expected for normal variables.
Therefore these results indicate that the return time series is consistent with the MDH.

\begin{figure}
\vspace{-4mm}
\centering
\includegraphics[height=4.2cm]{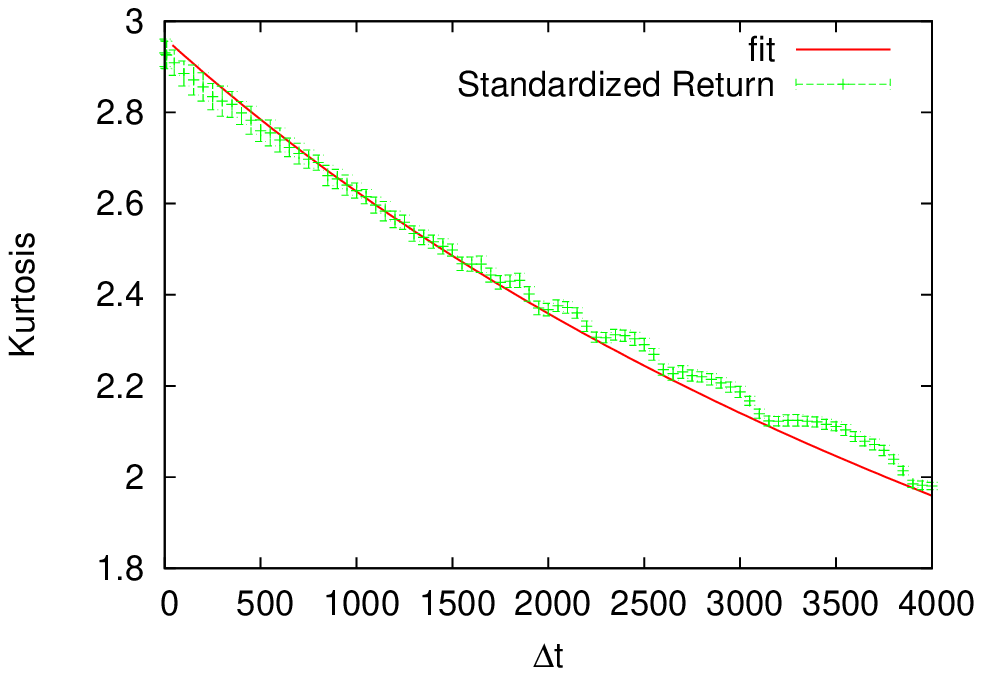}
\includegraphics[height=4.2cm]{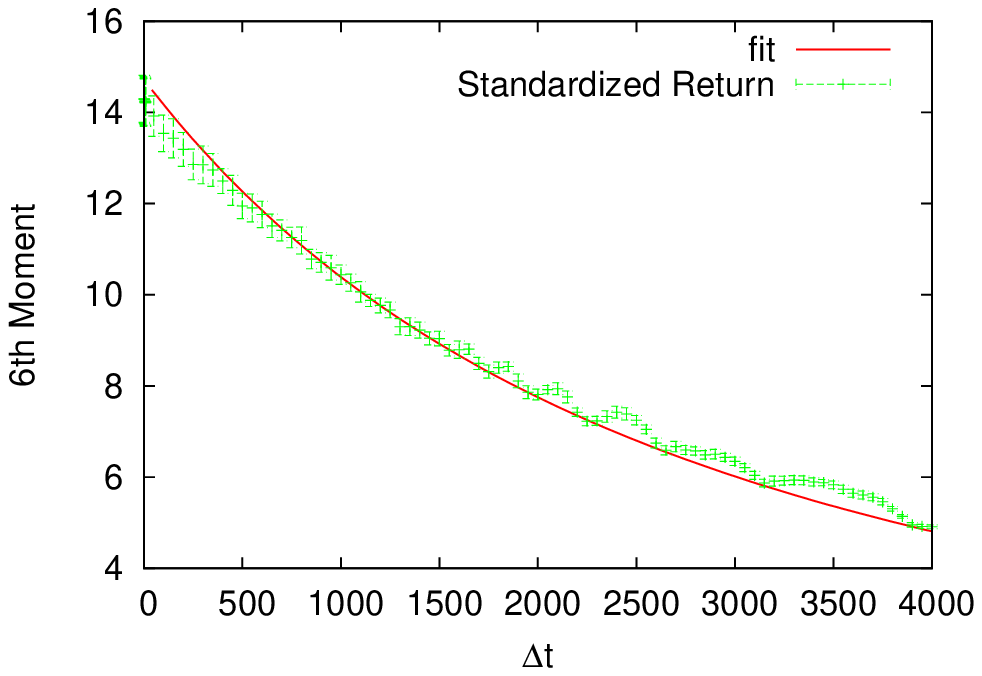}
\includegraphics[height=4.2cm]{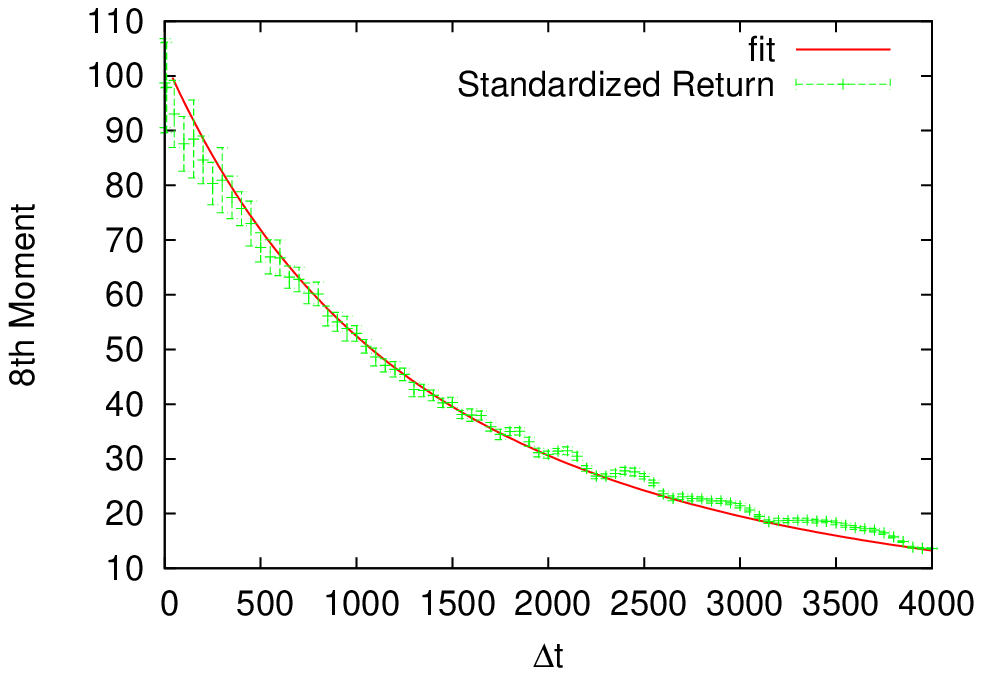}
\includegraphics[height=4.2cm]{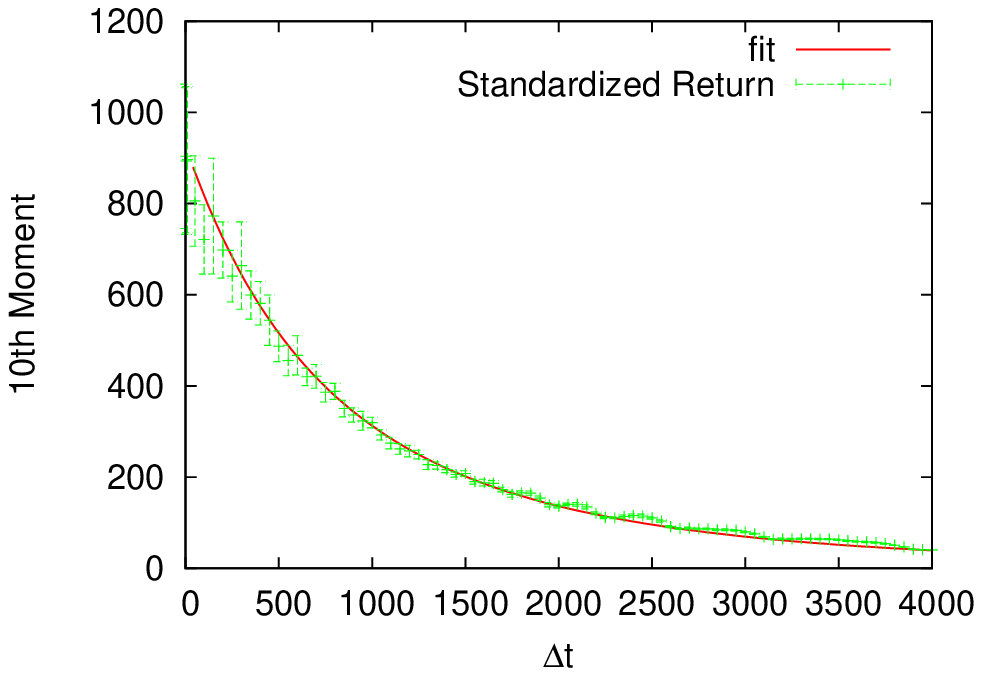}
\vspace{-1mm}
\caption{
Kurtosis, 6th, 8th and 10th moments of the SR as a function of  $\Delta t$.
The fittings are done by using the data in $\Delta t=[1,2000]$.
}
\vspace{-4mm}
\label{fig:Moment}
\end{figure}

\begin{table}[tbh]
\vspace{-2mm}
\centering
\caption{Theoretical values for standard normal variables and fitting results from the spin model. 
For the variance which has no finite-sample effect, the result at $\Delta t=1$ is given.}
\label{t1}
\begin{tabular}{c|ccccc}
\hline
 &variance & kurtosis&  6th&  8th&  10th \\
\hline  
 theory & 1  & 3 & 15 & 105 & 945 \\
 spin model & 1.002(9) & 2.96(3) & 14.72(4) &102.8(4) & 926(5) \\ 
\hline
\end{tabular}
\vspace{-8mm}
\end{table}

\section{Conclusion}
We have simulated a spin financial market 
and calculated the RV to investigate the return dynamics of the spin financial market.
The moments of the return standardized by the RV are strongly dependent on
the sampling interval except for variance. 
However we found that 
those moments converge to the expected theoretical values for normal variables
in the limit of $\Delta t \rightarrow 0$.
If the estimated volatility  is not precise enough the SR do not show
the exact normality. In \cite{Takaishi2} volatility of the spin model is estimated by
the GARCH model\cite{GARCH} 
and the kurtosis of the return standardized by GARCH volatility is found to deviate from 3,
which means the GARCH volatility may not be  precise enough. 
Our findings indicate that 
volatility of the returns is correctly calculated by the RV and the return dynamics 
of the spin financial market is consistent with the view of the MDH, 
i.e. the return time series follow a Gaussian process with time-varying volatility.

\section*{Acknowledgement}
\vspace{-1mm}
Numerical calculations in this work were carried out at the
Yukawa Institute Computer Facility.
This work was supported by JSPS KAKENHI Grant Number 25330047.


\begin{thebibliography}{4}

\bibitem{CONT}
R~.Cont: Quantitative Finance \textbf{1} (2001) 223.


\bibitem{Clark}
P.K.Clark:
Econometrica \textbf{41} (1973) 135.

\bibitem{Super}
T.Takaishi: Evol. Inst. Econ. Rev. 7(1) (2010)  89.



\bibitem{RV} 
T.G. Andersen and T. Bollerslev:
International Economic Review \textbf{39} (1998) 885.


\bibitem{RV3} 
T.G. Andersen, T. Bollerslev, F.X. Diebold and H. Ebens:
Journal of Financial Economics \textbf{61} (2001) 43.

\bibitem{RV4}
T.G. Andersen, T. Bollerslev, F.X. Diebold and P. Labys:
Multinational Finance Journal \textbf{4} (2000) 159.

\bibitem{RV6}
T.Takaishi, T.T.Chen, and Z.Zhen : 
Prog. Theor. Phys. Suppl. \textbf{194} (2012) 43.

\bibitem{Takaishi}
T.Takaishi:
Procedia - Social and Behavioral Sciences \textbf{65} (2012) 968.

\bibitem{Born}
S.Bornholdt:
Int. J. Mod. Phys. C \textbf{12} (2001) 667.

\bibitem{POTTS}
T.Takaishi:
Int. J. Mod. Phys. C \textbf{16} (2005) 1311.

\bibitem{Born2}
T.Kaizoji, S.Bornholdt and Y.Fujiwara:
Physica A \textbf{316} (2002) 441.

\bibitem{Born3}
S.M. Krause and S. Bornholdt:
(2011) arXiv:1103.5345.

\bibitem{Finite} 
R.T.Peters and R.G. De Vilder:
Journal of Business \& Economic Statistics \textbf{24} (2006) 444.


\bibitem{Andersen1}
T.G. Andersen, T. Bollerslev and D. Dobrev:
Journal of Econometrics \textbf{138} (2007) 125.

\bibitem{Andersen2}
T.G.Andersen, T. Bollerslev, P. Frederiksen and M.O. Nielsen:
J. Appl. Econometr. \textbf{25} (2010) 233.


\bibitem{Takaishi2}
T.Takaishi:
J. Phys.: Conf. Ser.  \textbf{454} (2013) 012041.

\bibitem{GARCH}
T.Bollerslev:
Journal of Econometrics \textbf{31} (1986) 307.


\end{thebibliography}
\end{document}